\documentclass{aa}
\usepackage{graphicx}
\usepackage{txfonts}
\usepackage{natbib}
\usepackage{url}

\newcommand{\bm}[1]{\mbox{\boldmath$#1$\unboldmath}}

\renewcommand{\arcsec}{.\hspace{-0.9mm}'\!\hskip0.4pt'\hspace{-0.2mm}}
\newcommand{\arsec}{'\!\hskip0.4pt'\hspace{-0.2mm}}

\begin{document}
   \title{Discriminant analysis of solar bright points and faculae
\thanks{Figures 11-14 are only available in electronic form via http://www.edpsciences.org}}

   \subtitle{I. Classification method and center-to-limb distribution}

   \author{P. Kobel \inst{1} \and J. Hirzberger \inst{1} \and S. K. Solanki \inst{1,2} \and A. Gandorfer \inst{1} \and V. Zakharov \inst{1} }

   \offprints{P. Kobel}

   \institute{Max-Planck Institut f\"{u}r Sonnensystemforschung, Max-Planck-Stra\ss e 2, 37191 Katlenburg-Lindau, Germany\\
              \email{kobel@mps.mpg.de}
              \and School of Space Research, Kyung Hee University, Yongin, Gyeonggi, 446-701, Korea}

   \date{}

%\abstract{}{}{}{}{}
% 5 {} token are mandatory

  \abstract
    {While photospheric magnetic elements appear mainly as Bright Points (BPs) at the disk center and as faculae near
the limb, high-resolution images reveal the coexistence of BPs and faculae over a range of heliocentric angles.
This is not explained by a ``hot wall'' effect through vertical flux tubes, and suggests that the transition from
BPs to faculae needs to be quantitatively investigated.}
    {To achieve this, we made the first recorded attempt to
discriminate BPs and faculae, using a statistical classification approach based on Linear Discriminant Analysis
(LDA). This paper gives a detailed description of our method, and shows its application on high-resolution images
of active regions to retrieve a center-to-limb distribution of BPs and faculae.}
     {Bright ``magnetic'' features were detected at various disk positions by a segmentation algorithm using
  simultaneous G-band and continuum information. By using a selected sample of those features to represent BPs
  and faculae, suitable photometric parameters were identified for their discrimination. We then carried out LDA
  to find a unique discriminant variable, defined as the linear combination of the parameters that best separates
  the BPs and faculae samples. By choosing an adequate threshold on that variable, the segmented features were
  finally classified as BPs and faculae at all the disk positions.}
    {We thus obtained a Center-to-Limb Variation (CLV) of the relative number of BPs and faculae, revealing the
  predominance of faculae at all disk positions except close to disk center ($\mu \geq 0.9$).}
% conclusions heading (optional),  leave it empty if necessary
  {Although the present dataset suffers from limited statistics, our results are consistent with other
  observations of BPs and faculae at various disk positions. The retrieved CLV indicates that at
  high resolution, faculae are an essential constituent of active regions all across the solar disk. We
  speculate that the faculae near disk center as well as the BPs away from disk center are associated with
  inclined fields.}

\keywords{Sun:photosphere - Sun:faculae, plages - Sun:magnetic fields - Methods:statistical - Techniques:high
angular resolution - Techniques:photometric}

   \titlerunning{Discriminant analysis of solar bright points and faculae I.}
   \authorrunning{P. Kobel et al.}
   \maketitle
%
%________________________________________________________________

\section{Introduction}

When imaged at high spatial resolution, the solar photosphere reveals a myriad of tiny bright features,
primarily concentrated in active regions and outlining the borders of supergranules in the quiet Sun. Near disk
center, they appear mainly as ``Bright Points'' (BPs) or ``filigree'' \citep{Dunn73, Mehltretter74}, i.e.
roundish or elongated bright features located in the intergranular downflow lanes \citep{Title87}, particularly
bright when observed in Fraunhofer's G-band \citep{Muller84, Berger95, Langhans02}. Near the limb, they
resemble more side-illuminated granules called ``faculae'' or ``facular grains'' \citep[e.g.][]{Muller75}, herein
considered as individual small-scale elements \citep{Hirz05}. The close association of BPs and faculae with
magnetic field indicators such as chromospheric \ion{Ca} {II} emission suggests that they are related phenomena
\citep{Mehltretter74, Wilson81}, both associated with small-scale kG flux concentrations \citep{Stenflo73}. These
so-called ``magnetic elements'' are considered as the basic building blocks of the photospheric magnetic activity
\citep[see][for reviews]{Schuessler_rev92, Solanki_rev93}, whence the importance of understanding their
fundamental physics. Also, much of the interest in faculae has been justified by their major role in producing
the total solar irradiance variation \citep{Lean88, Fligge88, Walton03, Krivova03}.

The peculiar appearance of BPs and faculae as well as their different distribution on the disk raises questions
about their physical origin and mutual relationship. The standard model accounting for both these phenomena
describes BPs and faculae as distinct radiative signatures of strongly evacuated thin flux tubes, arising from
different viewing angles \citep[``hot-wall'' model,][]{Spruit76, Knoelker88, Knoelker91, Steiner05}. This
simplified picture has been verified in its salient points by recent comprehensive 3D MHD simulations
\citep{Voegler05}. A major success has been the ability to qualitatively reproduce BPs near disk center
\citep{Schuessler03, Shelyag04} and faculae closer to the limb \citep{Keller04, Carlsson04}, thereby confirming
the basic hot-wall model to first order. However, images with the highest spatial resolution reveal the presence
of BPs away from the disk center, and of facular elements even close to the disk center \citep{Hirz05, Berger07}. Such
mixtures of BPs and faculae at several heliocentric positions is not explained by the hot-wall picture
considering vertical flux tubes, and seems not apparent in the simulated synthetic images \citep{Keller04}. Further, it is not
clear either whether the BPs and faculae seen at different heliocentric angles are associated with similar
magnetic structures, or rather with different structures prone to selection effects \citep{Lites04, Solanki_rev06}.
This shows that the transition from BPs to faculae is not clearly understood, and current models aiming
at reproducing BPs and faculae would benefit from a quantitative study of the distribution of these features on
the disk.

To tackle these issues, a necessary step is to sort the BPs and faculae observed at various disk positions, in
order to treat them separately. The approach proposed here is the first attempt in this direction, and relies on
Linear Discriminant Analysis (LDA) \citep{Fischer36} as a basis to ``classify'' features as BPs or faculae. Our
method makes use of purely photometric information, so that it only distinguishes the features
\emph{appearing} as BPs or as faculae. We applied this method to high-resolution images of active regions,
covering a range of heliocentric angles where the transition from BPs to faculae is expected. This allowed us to
retrieve, for the first time, an estimate of the center-to-limb variation of the relative amount of both
features, and thereby to quantitatively grasp how the appearance of magnetic elements varies from center to limb.

Although Discriminant Analysis has been fruitfully used in Astronomy \citep[see the general
review by][]{Heck89}, its application in the framework of solar physics thus far has been restricted to the study
of the conditions triggering solar flares \citep[the aim of ``probabilistic flare forecasting''][]{Smith96,
Leka03, Barnes07}, and to the response at the Earth's surface to the solar cycle \citep{Tung08}. Therefore, this
paper is intended to give a detailed description of our classification method, and by the same token provides a
concrete example of linear discriminant analysis applied to solar data. Among other potential applications in solar
physics, we mention the taxonomy of flares and the separation of chromospheric BPs and cosmic ray spikes on
wavelet-analyzed images \citep{Antoine02}.

The structure of this paper reflects the path taken to resolve the classification problem. Section
\ref{sec_proc&seg} describes the original dataset processing, and the automated segmentation method by which
bright features were detected at each disk position. Section \ref{sec_DA} outlines the classification scheme
while briefly presenting the principles of LDA. It also gives a detailed report of how this technique can be
applied to a selected sample of BPs and faculae in order to derive a single discriminant variable, based on which
a simple classification rule can be built. Section \ref{sec_classification} then deals with the actual
classification of all the segmented features, as well as the discussion of these results from a methodological
and physical point of view. Finally, Sect. \ref{sec_summary} summarizes the obtained results and gives future
directions for such work.

%__________________________________________________________________

\section{Image processing and segmentation}\label{sec_proc&seg}

\subsection{Dataset processing}\label{sec_processing}

The original dataset consists of simultaneous G-band (430.5 $\pm$ 0.5 nm) and nearby continuum (436.3 $\pm$ 0.5
nm) images recorded at the 1m Swedish Solar Telescope (SST, La Palma), on 7th and 8th September 2004. They cover
active regions at seven disk positions in the range $0.56 \leq \left< \mu \right> \leq 0.97$, where $\mu \equiv$
cos$\theta$, $\theta$ is the heliocentric angle and $\left< \mu \right>$ corresponds to the center of the
respective field of view (FOV), equivalent to the mean $\mu$ over the whole FOV (cf. Table \ref{table_data}).
This range of disk positions contains both BPs and faculae, and is thus well-suited to investigate their
transition. Because our study requires the highest spatial resolution in order to resolve individual BPs and
faculae, the dataset was restricted to the one to three best image pairs at each disk position (obtained
at peaks of seeing), which were kept for further processing and analysis (see Table \ref{table_data}).
\begin{table*}
\caption{Dataset specifications:
$(\mu_{\rm min}, \mu_{\rm max})$ indicates the $\mu$ coverage of the images, after their rotation along the disk
radius vector pointing towards the closest limb. FOV$_{\rm eff}$ is the effective field of view once the spots and pores have been masked
out. The number of image pairs selected for processing and analysis are given in the last column.}
\label{table_data}
\centering
\begin{tabular}{c c c c c c}     % 7 columns
\hline\hline
                      % To combine 4 columns into a single one
Date & NOAA & $\left< \mu \right>$ & $(\mu_{\rm min}, \mu_{\rm max})$ & FOV$_{\rm eff}$ [arcsec$^2$] & Number of Pairs \\
\hline
   07-Sept-2004 & 0669 & 0.97 $\pm$ 0.003 & (0.963, 0.977) & 2034 & 1 \\
   07-Sept-2004 & 0671 & 0.78 $\pm$ 0.008 & (0.75, 0.8) & 2135 & 1 \\
\hline
   08-Sept-2004 & 0670 & 0.97 $\pm$ 0.003 & (0.963, 0.976) & 1906 & 1 \\
   08-Sept-2004 & 0667  & 0.937 $\pm$ 0.003 & (0.928, 0.945) & 1027 & 2 \\
   08-Sept-2004 & -- & 0.9 $\pm$ 0.005 & (0.882, 0.916) & 2400 & 1 \\
   08-Sept-2004 & 0671 & 0.63 $\pm$ 0.01 & (0.58, 0.67) & 1923 & 3 \\
   08-Sept-2004 & 0671 & 0.6 $\pm$ 0.01 & (0.55, 0.64) & 1763 & 2 \\
   08-Sept-2004 & 0671 & 0.56 $\pm$ 0.01 & (0.51, 0.6) & 1904 & 1 \\
\hline
   13-Aug-2006 & 0671 & 0.77 $\pm$ 0.008 & (0.763, 0.776) & 566 & 1 \\
\hline
\end{tabular}
\end{table*}

\begin{figure*}
\centering
\includegraphics[width=\textwidth]{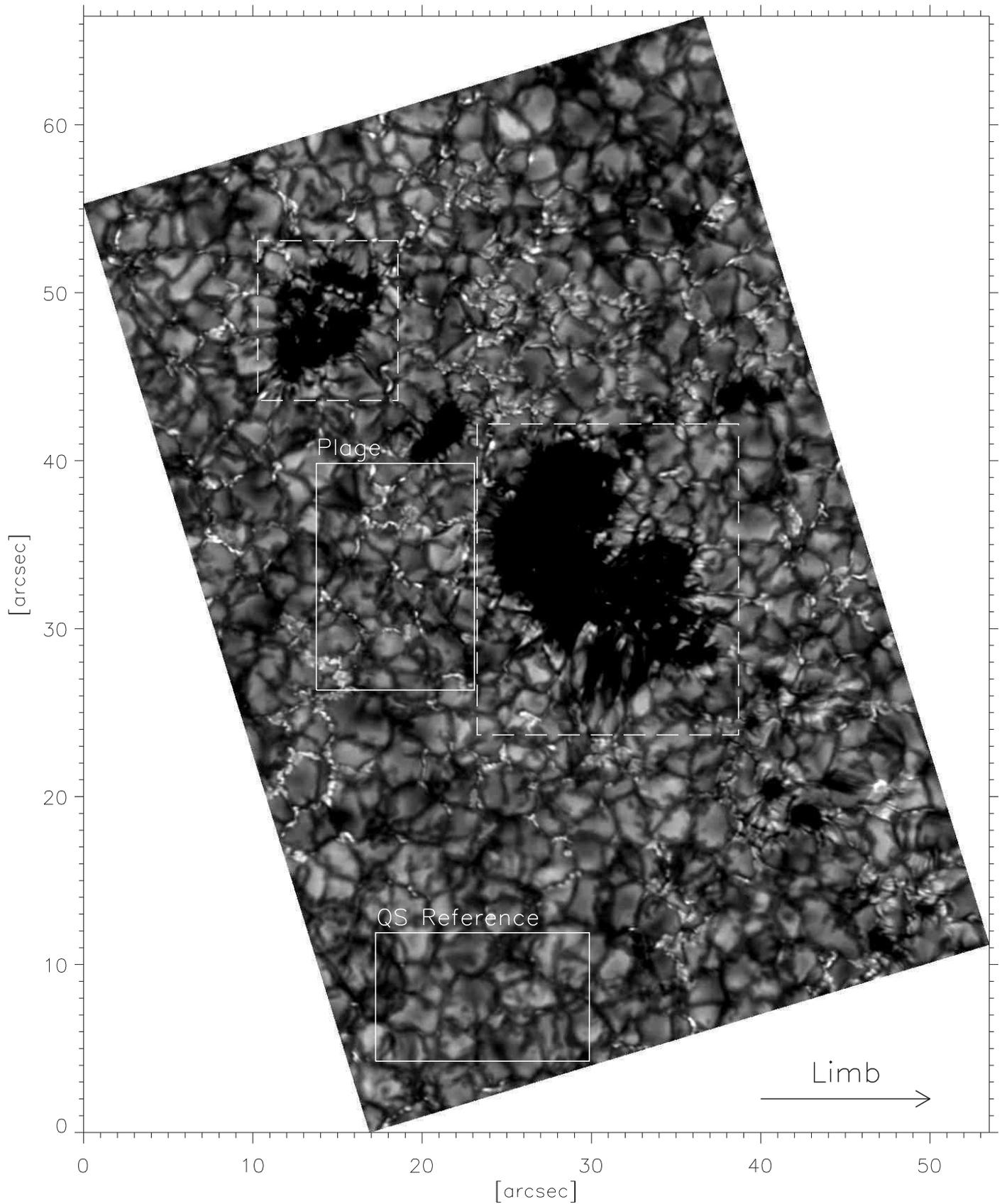}
\caption{G-band image of NOAA 0669 at $\left< \mu \right>$ = 0.97 recorded on 7th September, 2004. The dashed
lines outline the spot and pore areas masked out for the segmentation. The ``QS Reference'' indicates the
quasi-quiet Sun subfield chosen as reference for the contrast. The ``Plage'' subfield is the one selected for the
$C_{\rm G}$ \textit{vs.} $C_{\rm C}$ diagram shown in Fig.\ref{fig_GvsC}. The arrow indicates the direction of
the closest limb.}
\label{fig_imagemu97}
\end{figure*}

\begin{figure*}
\centering
\includegraphics[width=\textwidth]{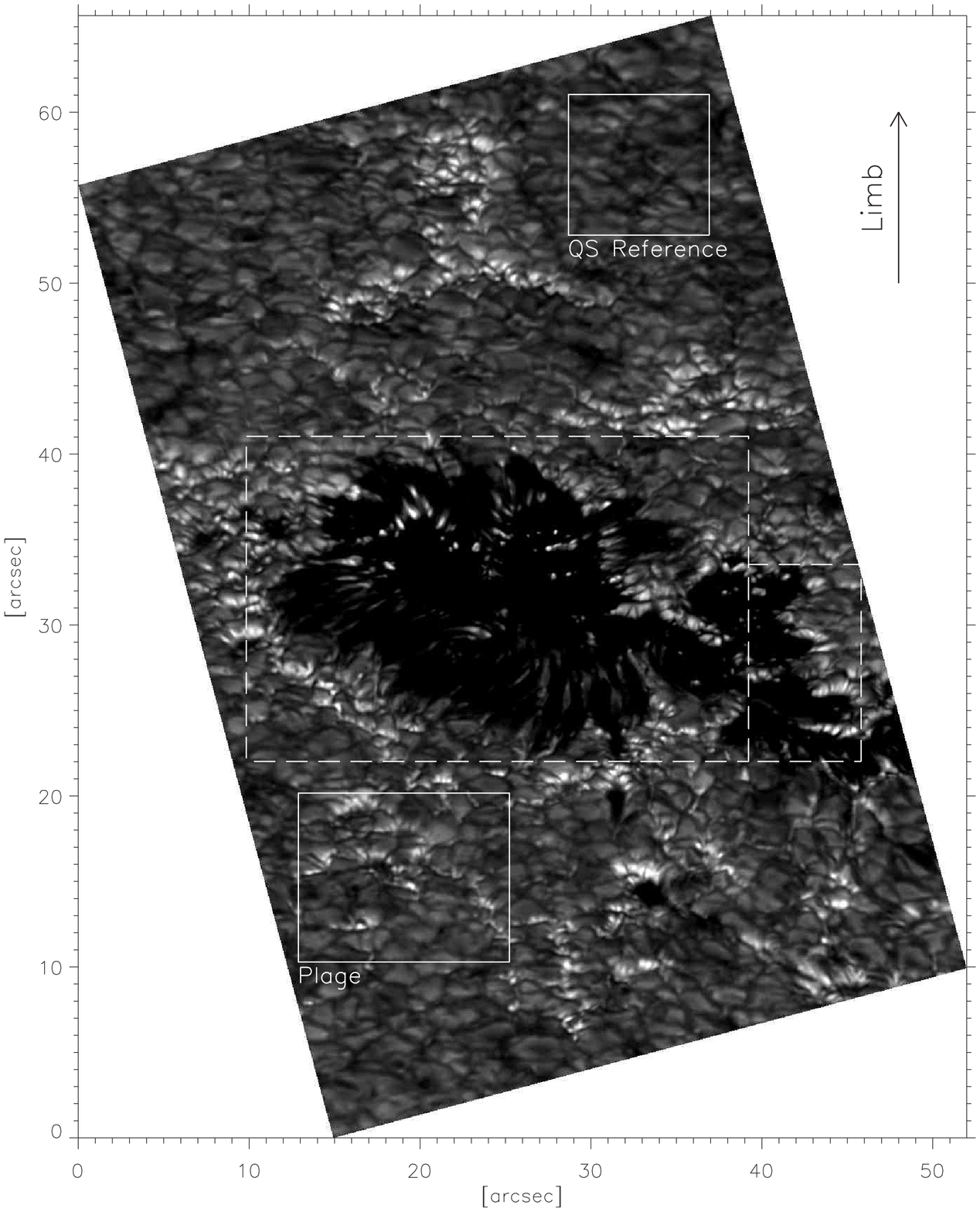}
\caption{G-band image of NOAA 0671 $\left< \mu \right>$ = 0.6 recorded on 8th September, 2004. The ``Plage''
area was used for the $C_{\rm G}$ \textit{vs.} $C_{\rm C}$ diagram plotted in Fig. \ref{fig_GvsC}.}
\label{fig_imagemu6}
\end{figure*}

For the selected image pairs, phase-diversity reconstruction allowed a roughly
constant angular resolution to be achieved, close to the diffraction limit ($\sim 0\arcsec 1$ at 430 nm). The reconstructed
simultaneous image pairs (G-band and continuum) were aligned and destretched using cross-correlation and grid
warping techniques (courtesy P. S\"{u}tterlin). The direction of the closest limb was found by comparison with
roughly co-temporal SOHO/MDI full disk continuum images, and the images were divided by the limb darkening
$\mu$-polynomial of \citet{Neckel94} at the nearest tabulated wavelength (427.9 nm). For each image pair, the
contrast $C$ was then defined relative to the mean intensity $\left< I \right>_{\rm QS}$ of a quasi-quiet Sun
subfield (of area ranging from 44 to 114 arcsec$^2$, depending on the image) as $C = (I - \left< I \right>_{\rm
QS})/\left< I \right>_{\rm QS}$. The G-band and continuum contrast are hereafter denoted $C_{\rm G}$ and $C_{\rm
C}$, respectively. To enhance the segmentation process (see Sect. \ref{sec_segmentation}), we applied a high-pass
spatial frequency filter to remove medium and large-scale fluctuations of the intensity (with observed spatial
scales between 5 and 30$\arsec$), presumably attributable to p-modes,
supergranular cell contrasts, straylight and residual flat-field effects. The Fourier filter was of the form
$f(k) = 1 - e^{-a^{2}k^{2}}$, where $k$ is the modulus of the spatial frequency, and the parameter $a$ was set to
have a cut-off frequency ($F(k) = 0.5$) of 0.2 arcsec$^{-1}$ and full power ($F(k) = 1$) at 0.65 arcsec$^{-1}$
\citep[in accordance with][]{Hirz05}. Finally, sunspots and large pores featuring umbral dots were masked out,
together with their immediate surrounding granules. This prevents the contamination of BPs/faculae statistics by
features of a different physical nature. Figures \ref{fig_imagemu97} and \ref{fig_imagemu6} show examples of
G-band images at $\left< \mu \right>$ = 0.97 and $\left< \mu \right>$ = 0.6, respectively, in which the quiet Sun
contrast reference and the masked out sunspot and pore areas are outlined.

\subsection{Magnetic brightening segmentation}\label{sec_segmentation}

Prior to their classification as BPs or faculae, bright magnetic features at the different disk positions
of our dataset were detected by a segmentation algorithm. The aims of our algorithm were twofold:
\begin{enumerate} \item{Detect magnetic brightenings photometrically by comparison of their contrast in G-band
and continuum.} \item{Decompose groups of BPs and striated faculae into individual elements by using
Multi-Level-Tracking \citep[MLT,][]{bovelet01,Bovelet03, Bovelet07}.} \end{enumerate} The second point
significantly increases the statistics of the study, and relies on the assumption that these elements correspond
to distinct magnetic features. This has been justified for intergranular BPs at disk center \citep{Berger01},
while observations of the dynamic behavior of striated faculae suggest a correspondence with those BPs
\citep{DePontieu06}, the dark striations being associated with sites of lower magnetic field strength
\citep{Berger07, Carlsson04}.

\begin{figure*}[bht]
\centering
\includegraphics[width=0.9\textwidth]{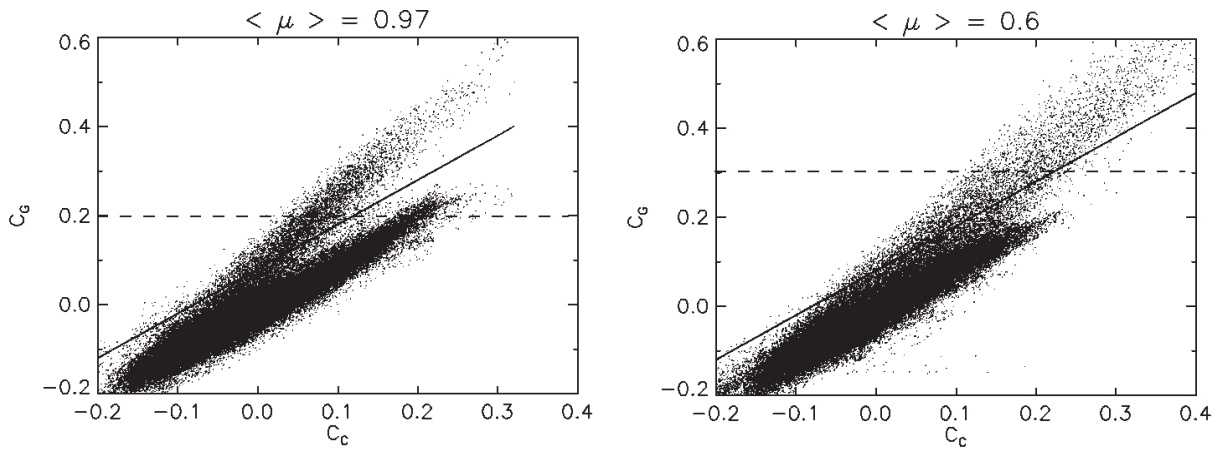}
\caption{$C_{\rm G}$ \textit{vs.} $C_{\rm C}$ scatterplots for selected plage subfields of area approximately $10
\times 12$ arcsecs$^2$ at $\left< \mu \right> = 0.97$ (\textit{left}) and $\left< \mu \right> = 0.6$ (\textit{right}). The exact
locations of these subfields within their respective images are outlined in Figs. \ref{fig_imagemu97} and
\ref{fig_imagemu6} (rectangles denoted ``Plage''). The solid line corresponds to the difference threshold $C_{\rm
diff,t}$ and the dashed line to the G-band threshold $C_{\rm G,t}$.}
\label{fig_GvsC}
\end{figure*}

The principle behind point 1 is best illustrated by $C_{\rm G}$ \textit{vs.} $C_{\rm C}$ scatterplots of
a plage area, as shown in Fig. \ref{fig_GvsC} (see Figs. \ref{fig_imagemu97} and \ref{fig_imagemu6} for the location
of the chosen plage subfields). As can be seen in Fig. \ref{fig_GvsC}, the scatterplot splits into two clearly
distinct pixel distributions. A similar pattern appears in the diagnostics of radiative MHD simulations of
\citet{Shelyag04}, where the upper distribution is shown to be associated with strong flux concentrations,
whereas the lower one corresponds to weakly magnetized granules \citep[see also][for the comparison of different
1D LTE atmospheres]{Sanchez01}. We can thus select pixels which are G-band bright and likely to be of magnetic
origin by imposing two thresholds: a G-band threshold $C_{\rm G,t}$ selecting the bright portion of the diagram
(dashed lines in Fig. \ref{fig_GvsC}), and a threshold $C_{\rm diff,t}$ on the contrast difference $C_{\rm diff}
\equiv (C_{\rm G} - C_{\rm C})$ \citep[see][for more details]{Berger98}.

To achieve point 2, we chose a set of closely-spaced MLT levels between $C_{\rm G,t}$ and $C_{\rm G} =
0.7$. The interlevel spacing was tuned to 0.02 \citep[similar to][for BPs at disk center]{Bovelet07} by visual
comparison of the segmentation maps with the original images. This spacing allowed chains of BPs and faculae
striations to be resolved, while avoiding over-segmentation. The structures were then extended down to $C_{\rm G}
= 0$ with two intermediate levels at $C_{\rm G} = 0.1$ and $C_{\rm G} = 0.05$. This extension increases the
segmented area of faculae compared to BPs, allowing its further use as discriminant parameter (see Sect.
\ref{sec_params}). The intermediate levels prevent the merging of BPs with adjacent granules and the clumping of
granular fragments when the contrast of intergranular lanes does not drop below $C_{\rm G} = 0$. Since a
necessary condition for a feature to be selected is to have its contrast maximum above $C_{\rm G,t}$, no other
levels were included between $C_{\rm G} = 0$ and $C_{\rm G,t}$ to avoid oversegmentation. Likewise, structures
of less than 5 pixels in area (corresponding to the area of a roundish feature with a diameter of $0\arcsec 1$,
i.e. roughly equal to the diffraction limit) were removed at each MLT level.

The segmentation algorithm then proceeded in two steps: First, MLT was applied to the spatially-filtered
G-band images. In a second step, structures corresponding to ``magnetic'' features were selected by requiring
them to \emph{contain} a minimum of 5 pixels satisfying $C_{\rm G} > C_{\rm G,t}$ and $C_{\rm diff} > C_{\rm
diff,t}$. A binary map of segmented features was ultimately obtained for each G-band/continuum image pair.

\begin{figure}
\centering
\includegraphics[width=\columnwidth]{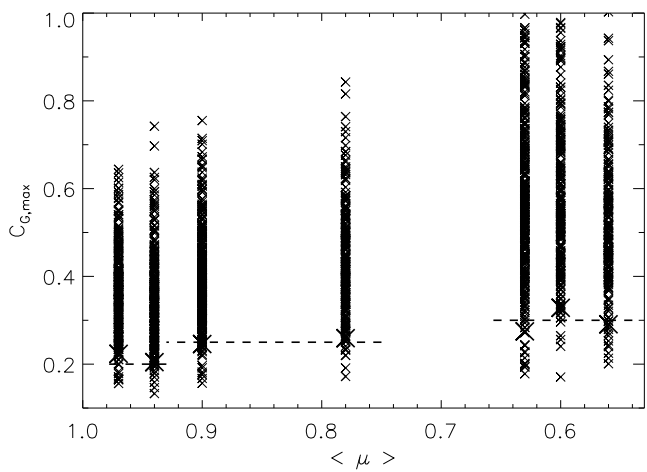}
\caption{Maximum G-band contrast values (small crosses) of all features segmented only with a difference
threshold $(C_{\rm diff,t} = 0.1)$ at each disk position. The dashed line corresponds to the chosen G-band threshold
removing approximately the $4 \%$ faintest features, which are delimited exactly by the large crosses.}
\label{fig_GtCLV}
\end{figure}

From Fig. \ref{fig_GvsC}, one notices that the ``magnetic'' and ``non-magnetic'' pixel distributions overlap more
at $\left< \mu \right> = 0.6$ than at $\left< \mu \right> = 0.97$, a tendency that was generally observed for
decreasing $\left< \mu \right>$. To avoid the false detection of granules, the G-band threshold must then be
raised as $\left< \mu \right>$ decreases, inasmuch as the ``magnetic'' pixel distribution extends towards
larger values of $C_{\rm G}$ while the ``non-magnetic'' one reaches lower values (as the rms contrast of granules
decreases towards the limb). To do this consistently, we determined a CLV of maximum G-band contrasts $C_{\rm
G,max}$ of features segmented \emph{independently} of the G-band threshold. Specifically, taking one image pair
at each disk position, the features were segmented solely by a safe difference threshold $C_{\rm diff,t} = 0.1$,
and the ones having fewer than 20 pixels above this threshold were removed \citep[as most non-magnetic detections
contain only a few pixels,][]{Berger01}. A visual count over a portion of the images yielded an estimate of the
remaining fraction of false detections, approximatively 4~\%. Under the reasonable assumption that these false
detections were also the faintest, we adjusted the G-band threshold consistently at all $\left< \mu \right>$
to eliminate the $\sim$ 4 \% faintest features (the chosen value of the threshold was actually rounded up, and
was constant throughout the field of view at each $\left< \mu \right>$). The values of the features maximum
G-band contrast $C_{\rm G,max}$ as well as the $\left< \mu \right>$-dependent G-band thresholds are plotted
\textit{vs.} $\left< \mu \right>$ in Fig. \ref{fig_GtCLV}.

Unlike the G-band threshold, the difference threshold $C_{\rm diff,t}$ can be set to a unique value for all disk
positions, inasmuch as the ``non-magnetic'' distribution has a slope roughly equal to unity at all
$\left< \mu \right>$. To set $C_{\rm diff,t}$ properly, we made use of ``test'' data consisting of a single
G-band/continuum image pair obtained with the same setup as our original dataset (and processed as in
Sect. \ref{sec_processing}, except for speckle reconstruction), but supplemented with SOUP (Lockheed Solar
Optical Universal Polarimeter) Stokes $V$ and $I$ maps, recorded in the wing of the \ion{Fe}{I} 6302.5 {\AA} line
with a detuning of 75 m{\AA}. Given the value of the G-band threshold for that disk position, the difference
threshold was tuned such as to minimize the fraction of ``false'' detections. By considering false detections as
having less than 5 pixels with $|V/I| \geq 0.075$ (well above the noise level $\sim$$10^{-2}$), the optimal
difference threshold was found as $C_{\rm diff,t} = 0.08$. The corresponding fraction of false detections amounts
to roughly 2~\%. These test images were, however, not used further because they were focused on a large sunspot
and hence contain a very small effective field of view (see Table \ref{table_data}, 13-Aug-2006).

Without information about the magnetic field itself, our segmentation has to rely on purely photometric
thresholds, and hence cannot detect \emph{all} the magnetic features. The combined thresholds only aim
at detecting a sample of bright features that is \emph{least biased} by non-magnetic ones. However, the use of
thresholds always implies the drawback of selection effects. In particular, the G-band threshold will neglect
fainter features, especially low-contrast BPs near disk center \citep[see][]{Title96, Bovelet07, Shelyag04}.

\section{Discriminant analysis of bright points and faculae}\label{sec_DA}

\subsection{General scheme and training set}\label{sec_scheme}

To develop an algorithmic classification method for BPs and faculae, we adopted the following scheme, that uses
Linear Discriminant Analysis \citep[LDA, a statistical technique first introduced by][]{Fischer36} on a reference
sample of features:
\begin{enumerate}

\item{\emph{Training set selection}: Extraction of a reference sample of features, visually identified as BPs and faculae.}

\item{ \emph{Discriminant parameter definition}: Choice of observables taking sufficiently different values for
the BPs and faculae of the training set, in order to be of use for the further discrimination of the rest of
features.}

\item {\emph{LDA}: Determination of a unique variable by linear combination of the chosen
parameters, such that it best discriminates between the two classes of the training set.}

\item{\emph{Assignment rule}: Imposition of an adequate threshold on the discriminant
variable defined by LDA, separating the BPs and the faculae of the training set.}

\end{enumerate}
Ultimately, \emph{all} the magnetic brightenings detected by the segmentation algorithm can be classified
according to the assignment rule, by measuring their value of the variable defined by LDA. Comprehensive
manuscripts about the general topic of classification can be found in \citet*{Murtagh87} and \citet{Hand81}.

Our training set was chosen as a sample of 200 BPs and 200 faculae, obtained by manual selection of 40 features
of each at each of five disk positions: $\{\left<\mu\right> = 0.56, 0.63, 0.78, 0.9, 0.97\}$ for faculae and
$\{\left<\mu\right> = 0.63, 0.78, 0.9, 0.94, 0.97\}$ for BPs. Because it is used as a reference for the classes,
the selected sample should be statistically representative of the actual populations of BPs and faculae (such as
would be identified by eye). At each disk position, care was thus taken to select the most homogeneous mixture of
features with various contrasts and sizes, distributed over the whole field of view.

It should be kept in mind that BPs and faculae are possibly not two distinct types of objects, but the radiative
signatures of more or less similar physical entities (magnetic flux concentrations) viewed under different
angles. Consequently, there may well be no sharp boundary between the two classes, but rather a continuous
transition with a spectrum of ``intermediate features'', having various degrees of ``projection'' onto the
adjacent limbward granules \citep[see][and Sect. \ref{sec_profiles}]{Hirz05}. The concept of classes can
nonetheless be introduced to represent the populations of features that would be reasonably identified as BPs and
faculae upon visual inspection, but the approach proposed here cannot claim to classify the intermediate features
mentioned above.

\subsection{Characteristic profiles}\label{sec_profiles}

As a basis to define discriminant parameters, we considered the spatial variation of G-band contrast along a cut
made through a BP or a facula. Small magnetic features are indeed known to exhibit more pronounced signatures in
the G-band than in continuum, and such contrast profiles have characteristic shapes when BPs and faculae are cut
along specific directions: radially for limb faculae, and across the intergranular lane for disk center BPs
\citep{Berger95, Hirz05}.

\begin{figure}[thb]
\centering
\includegraphics[width=\columnwidth]{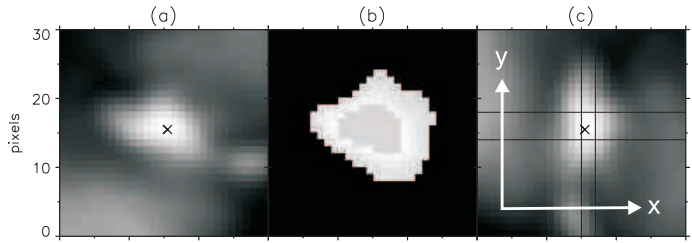}
\caption{Orientation of an individual feature in its local $x/y$ coordinate frame. \textbf{a)} Zoom window surrounding
the feature in the original G-band image. The cross indicates the location of the contrast maximum. \textbf{b)} Isolated
feature as delimited by the segmentation map. The pixels having $C_{\rm G} \geq 0.5$ $C_{\rm G,max}$ are
highlighted in grey, and are used to compute the G-band contrast moment of inertia. \textbf{c)} Window rotated such that
the $y$-component of the G-band contrast moment of inertia $ M_{\rm G,y}$ is minimum, thereby defining the local
$x/y$ coordinate frame. The rows and columns used to retrieve the average profiles along $x$ and $y$ are
contained between the straight black lines.}
\label{fig_orientation}
\end{figure}

\begin{figure*}[hbt]
\centering
\includegraphics[width=0.9\textwidth]{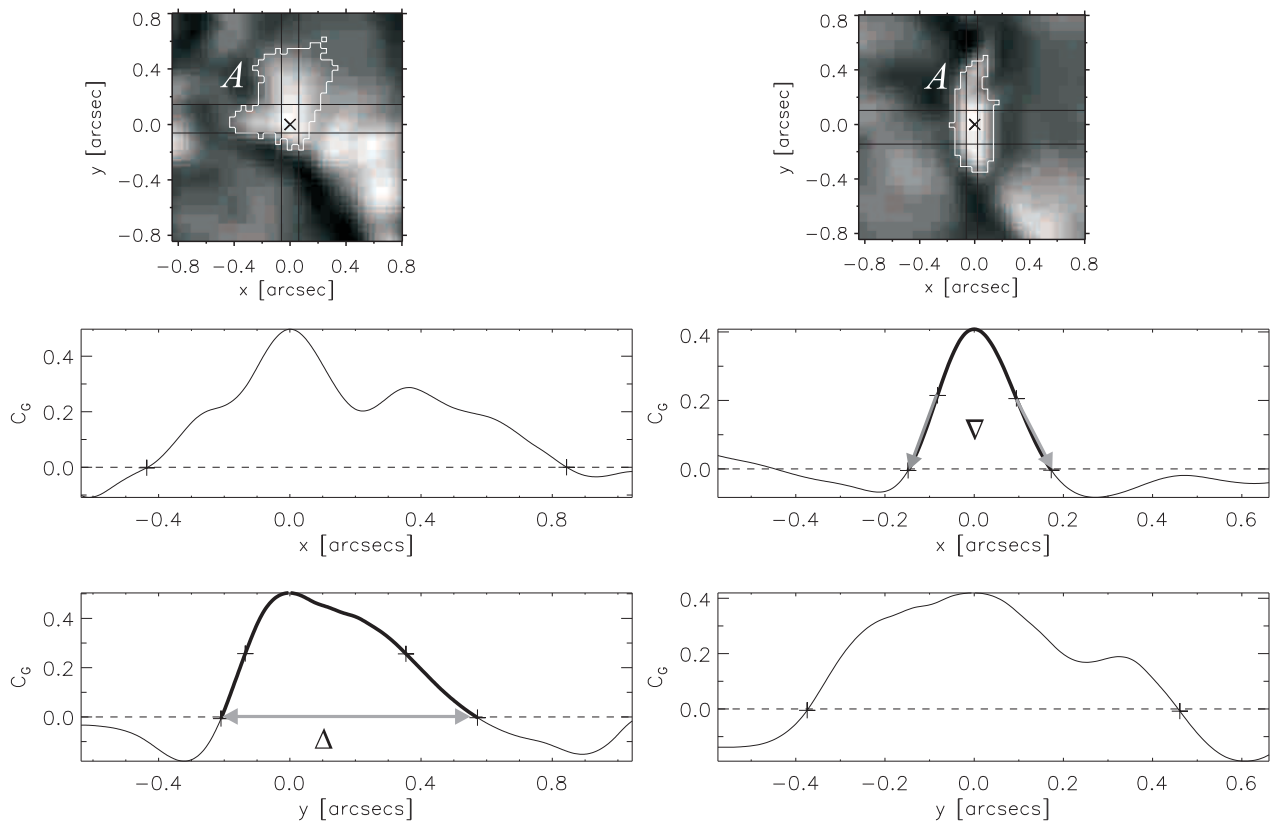}
\caption{Local frame orientation and G-band contrast profiles and of a typical facula (\textit{left}) and a
typical BP (\textit{right}). \textit{Top windows}: Orientation of the features in their local $x/y$ coordinate
frames, where the black lines delimit the pixels having $C_{\rm G} \geq 0.9 C_{\rm G,max}$ used for profile
averaging. The white contours obtained from the segmentation map enclose the area $\textit{A}$ of the features.
\textit{Lower panels}: Average G-band contrast profiles along $x$ and $y$. The retrieved \emph{characteristic
profile} for the BP and the facula is indicated by the thick lines. The ``+'' marks intersecting the reference
level $C_{\rm G} = 0$ (dashed line) delimit the positive contrast portion of the profiles, and the upper ``+''
marks indicate the half-max level $C_G = 0.5 C_{G,max}$ on the characteristic profiles. The parameters $\Delta$
and $\nabla$ are illustrated on the characteristic profiles of the facula and the BP, respectively. All the $x/y$
profiles were cubic spline-interpolated by a factor of 10, in order to avoid artificial roughness due to
sampling when choosing the characteristic profile (as the smoothest of the $x/y$ profiles, see Sect.
\ref{sec_profiles}).}
\label{fig_profiles}
\end{figure*}

The following procedure was developed to retrieve \emph{one} characteristic profile per
feature, independently of the feature type and disk position: First, each feature was oriented in a local
coordinate frame $x/y$ as illustrated in Fig. \ref{fig_orientation}. The $x/y$ axes were defined such as to
minimize the $y$-component of the feature's G-band ``contrast moment of inertia'' $ M_{\rm G,y} \equiv \sum
C_{\rm G}(x,y)(x-x_{\rm max})^2$, where $x_{\rm max}$ is the $x$-location of the contrast maximum $C_{\rm
G,max}$. To give optimal results on the orientation, the summation only ran over pixels having $C_{\rm G} \geq
0.5$ $C_{\rm G,max}$, thus involving only the ``core pixels'' of the features\footnote{Involving pixels
with lower contrast yields poorer results, as these are often associated with granulation in the case of faculae,
and thus do not carry information about the orientation of the facular brightening itself.}. In practice, the
minimum of $M_{\rm G,y}$ was found by iteratively rotating a small window surrounding the feature with $5^\circ$
steps (smaller steps did not yield better results, due to the finite number of pixels considered). Next, contrast
profiles were obtained along $x$ and $y$ by averaging the rows and columns of that window having pixels with
$C_{\rm G} \geq 0.9$ $C_{\rm G,max}$ (delimited by black lines in Fig. \ref{fig_orientation}c). Such $x/y$
profiles are displayed in Fig. \ref{fig_profiles} in the case of a typical BP (\textit{right}) and a typical
facula (\textit{left}). These profiles were further restricted to the contrast range $C_{\rm G} > 0$ about
$C_{\rm G,max}$ (delimited by the lower ``+'' marks), such that all profiles share a consistently-defined
reference level $C_{\rm G} = 0$. Finally, the single \emph{characteristic profile} for each feature was found to
be the smoothest of the positive contrast-restricted $x/y$ profiles (overplotted in thick). To quantify the
smoothness of the profiles, we counted the number of their local extrema, eventually adding the number of
inflexions if the number of extrema was equal in $x$ and $y$. The use
of MLT segmentation (as opposed to a single-clip) is an essential prerequesite for obtaining these
characteristic profiles, by avoiding that pixels from adjacent features contaminate the contrast moment of
inertia and thus spoil the feature's orientation process.

Owing to the previous orientation of the features, the characteristic
profiles exhibit different shapes for BPs and faculae, and consequently proved very useful for the extraction of
valuable discriminant parameters (see Sect. \ref{sec_params}). In contrast, profiles retrieved along the
disk radius vector (as performed in early stages of this work) have less characteristic shapes and thus less
power to distinguish BPs from faculae, due to the scatter in the orientation of these features with respect to the
radial direction. As can be seen in the examples of Fig. \ref{fig_profiles}, the characteristic profile of the
typical BP is narrower and steeper than the profile of the typical facula. In particular, the characteristic
profile of the facula is indistinguishable from the adjacent granule, as the contrast varies monotonously from
one to the other. We mention the resemblance of the characteristic profiles of the BP and facula to the
observations of \citet{Berger95} and \citet{Hirz05}, respectively, as well as with the synthetic profiles of
\citet{Knoelker88} and \citet{Steiner05}.

\begin{figure}[!htb]
\centering
\includegraphics[width=0.9\columnwidth]{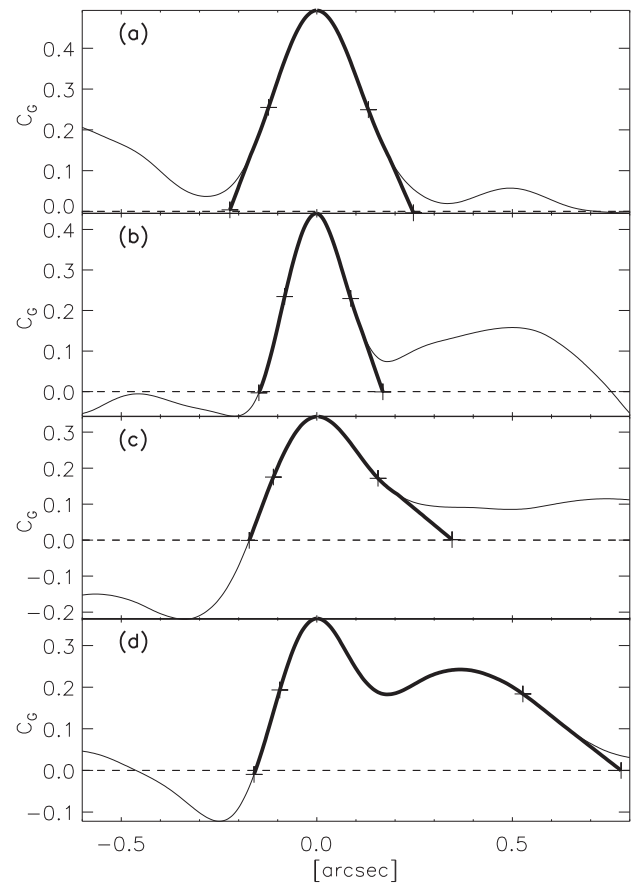}
         \caption{Characteristic profiles (thick lines) of BPs \textbf{(a, b)}
         and intermediate features \textbf{(c, d)} surrounded by one or two
         ``grey'' lanes. On the grey lane sides, the characteristic
         profiles have been linearly extrapolated from the half-max
         level (``+'') to $C_{\rm G} = 0$. From \textbf{a)} to \textbf{d)}, these
         profiles illustrate the continuous transition between the
         typical BPs and typical faculae, for which examples are shown in Fig. \ref{fig_profiles}.}
\label{fig_intermediate}
\end{figure}

Due to finite resolution, straylight, and the partial compensation of spatial intensity fluctuations by the
filter (see section \ref{sec_processing}), it is common to find BPs embedded in ``grey'' lanes with positive
contrast \citep{Bovelet07}. Upon careful visual analysis of grey lane-BPs profiles, we identified these grey
lanes as contrast depressions with a low minimum ($C_{\rm G,min} \leq 0.1$), separating the BP profile from the
adjacent granule profile. As most normal BPs profiles have quasi-linear slopes at their edges, the sides of
profiles featuring grey lanes were linearly extrapolated down to $C_{\rm G} = 0$. Only after this could the $x$
and $y$ average profile be properly restricted to positive contrast values, and their smoothness compared for the
adequate retrieval of the characteristic profile. This linear extrapolation is illustrated in Fig.
\ref{fig_intermediate} for characteristic profiles of both BPs and ``intermediate features'', indicating at the
same time the variety of feature profiles that can be obtained.

\subsection{Discriminant parameters}\label{sec_params}

In search of adequate discriminant parameters, we carried out a pilot study by defining a set of
parameters. These included the peak-to-width ratio, area asymmetry and second moment of the characteristic
profiles, the local contrast relative to the immediate surroundings \citep[similar to][]{Bovelet03}, and the
contrast of adjacent lanes. By looking at the distribution of the parameter values for the BPs and faculae of the
training set (mean values and standard deviation at each $\left<\mu\right>$, see below) as well as their
correlation, three roughly mutually independent parameters were eventually found to be good discriminants for the
training set classes. Their definitions are illustrated in Fig. \ref{fig_profiles}:
\begin{itemize}
\item $\Delta$ := width of the characteristic profile at the reference level $C_{\rm G} = 0$ [arcsecs].
\item $\nabla$  := average slope (from both sides) of the characteristic profile below the half-max level $C_{\rm G,HM} = 0.5C_{\rm G,max}$ [arcsecs$^{-1}$].
\item $\textit{A}$ := apparent area (projected onto the plane of the sky) of the feature defined by the segmentation binary map [arcsecs$^2$].
\end{itemize}
We emphasize that the three chosen parameters are defined using \emph{relative} contrast levels ($C_{\rm G} = 0$
and $C_{\rm G} = 0.5C_{\rm G,max}$), allowing the comparison and classification of features having
different absolute contrast values (notably due to the CLV of contrast).

\begin{figure*}
\centering
\includegraphics[width=0.9\textwidth]{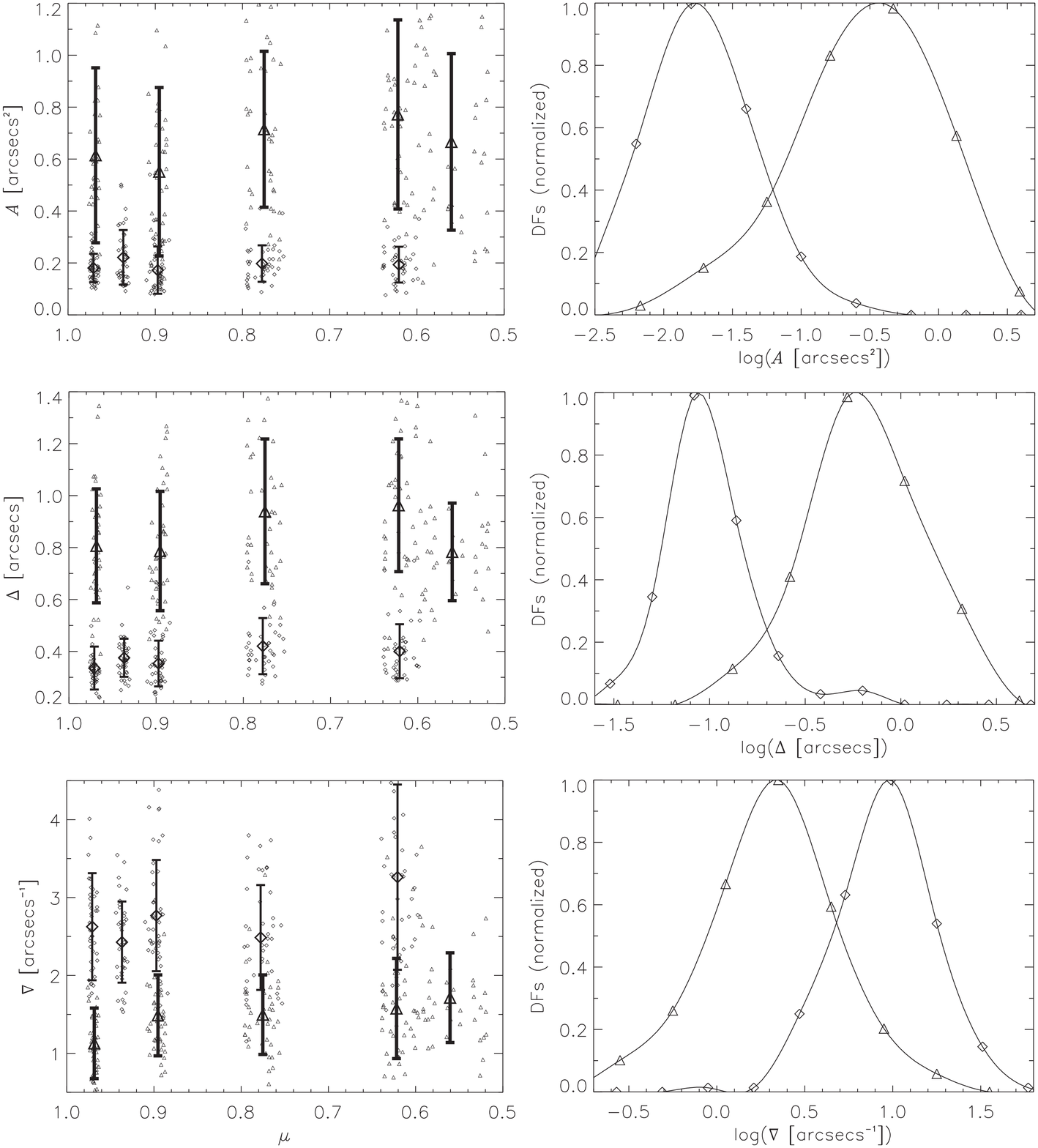}
\caption{\textit{Left column}: Mean values and standard deviations of the three parameters for the BPs
(``$\diamond$'' and thin bars) and faculae (``$\triangle$'' and thick bars) of the training set \textit{vs.} disk
position $\left<\mu\right>$. The training set does not contain BPs at $\left<\mu\right> = 0.56$ or faculae
at $\left<\mu\right> = 0.94$. To give an idea of the outliers and the $\mu$ distribution of the training set
features at each disk position, individual features values are overdrawn (small ``$\diamond$'' for BPs and
small ``$\triangle$'' for faculae). The $\mu$ values of the individual features were computed by using the
$\left<\mu\right>$ (center of FOV) of the corresponding images as reference. \textit{Right column}: Normalized
density functions (DFs) histograms of the log-transformed parameters for BPs (``$\diamond$'') and faculae
(``$\triangle$''), obtained by combining all disk positions of the training set together. Cubic splines are
overplotted for clarity and their maxima were used for the normalization of the histograms.}
\label{fig_params}
\end{figure*}

Fig. \ref{fig_params} (left column) shows the mean values and standard deviations of the parameters $\textit{A}$,
$\Delta$ and $\nabla$ at the $\mu$-values of the training set. These parameters describe well the
different appearances of BPs and faculae for the following reasons. The best discriminant parameter,
$\Delta$, takes greater values for faculae as it encompasses the width of the adjacent granular profile (as the
facular and granular profile are merged together, cf. Fig. \ref{fig_profiles}), whereas BPs are limited to the
width of intergranular lanes. The parameter $\nabla$ describes how steeply the contrast drops towards the edges of
the profile and typically has larger values for BPs, which show steep and symmetric contrast enhancements
squeezed between the adjacent granules. To supplement these two profile parameters, the segmented feature area
$\textit{A}$ has been added to avoid that faculae with small widths (typically lying on small abnormal granules
frequently found in active regions) are classified as BPs. In area these faculae appear significantly larger.

As can be seen from Fig. \ref{fig_params}, the parameter values do not vary significantly with $\left< \mu
\right>$, as the difference between the largest and smallest mean values over the whole $\mu$ range barely exceed
their standard deviations. The relative constancy of width and area is particularly surprising for
faculae, and could be due to a compensation of granular foreshortening by enhanced radiative escape in the
direction toward the flux concentration \citep{Steiner05}, as well as to the distribution in the orientations of
faculae (to be discussed in a forthcoming paper). The relative invariance of the parameters is nevertheless
advantageous, as it justifies performing LDA on the whole training set at once (all $\left< \mu \right>$
together), thus allowing us to find a single linear combination of parameters and a single BPs/faculae threshold
valid for all the disk positions of our dataset. Moreover, combining all the training set features enhances the sampling of the classes and yields a more accurate threshold.

It should be noted that the values of the parameters $\textit{A}$, $\Delta$ and $\nabla$ depend to some extent on
spatial resolution. This points to the necessity of having a dataset of roughly constant resolution, a condition
met by our selected image pairs (see Sect. \ref{sec_processing}).

\subsection{Linear Discriminant Analysis}\label{sec_LDA}

Because LDA distinguishes classes based solely on means and covariances (see Equation \ref{eqn_J}), it works best
for parameters that are normally or at least symmetrically distributed \citep{Murtagh87}. To verify this
condition, we studied the density functions (DFs) of our three parameters by producing histograms of the training
set, and estimated the skewnesses via the third standardized moment \citep{Kenney62}. Taking the natural
logarithm was found to reduce the skewness of all parameters \citep{Stahel01}, and therefore they were replaced
by their log-transforms\footnote{This was partly expected, as the width of magnetic bright points has
been observed to be log-normally distributed \citep{Berger95}.}. The DFs of
log($\textit{A}$), log($\Delta$) and log($\nabla$) are displayed in Fig. \ref{fig_params} (right column).

\begin{figure*}
\centering
\includegraphics[width=0.9\textwidth]{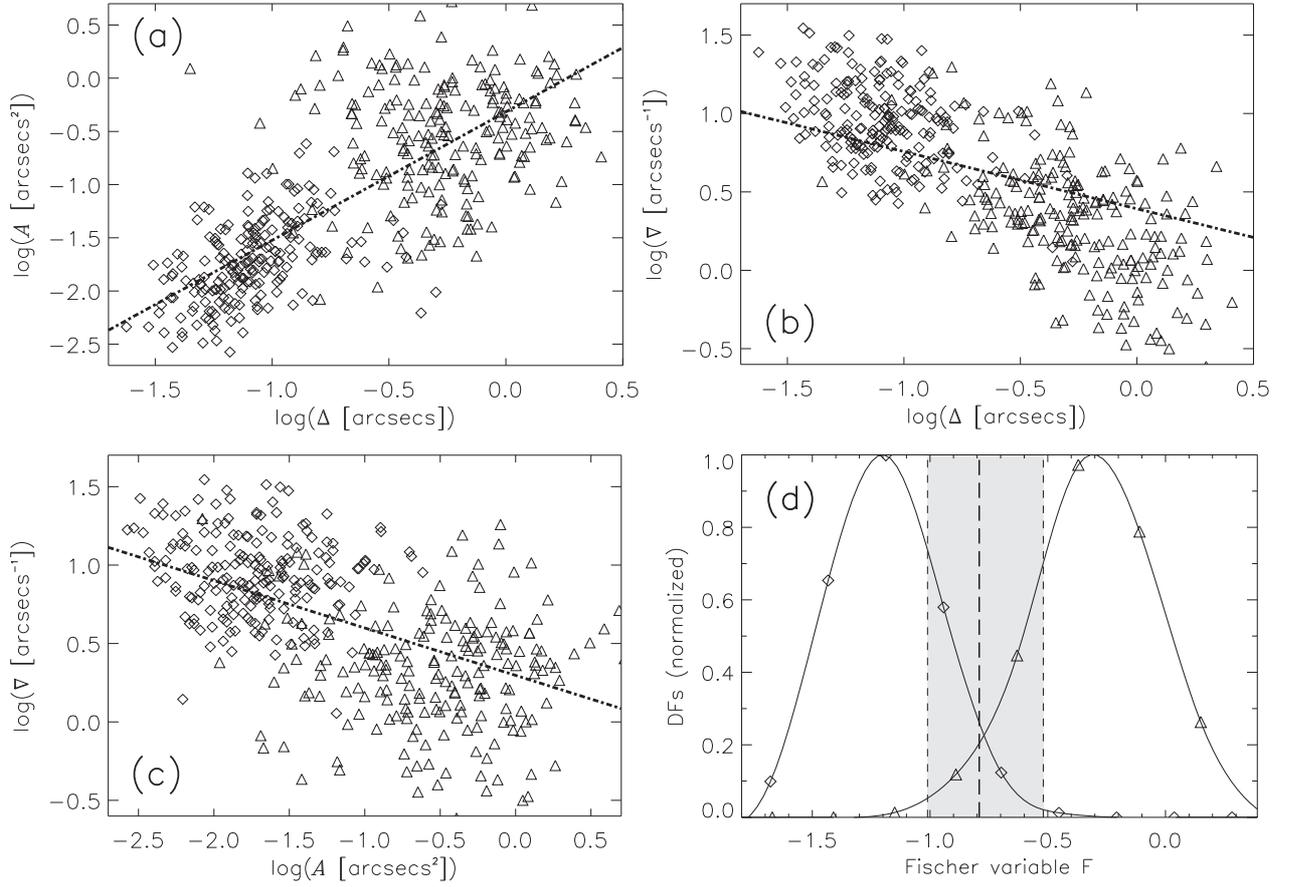}
\caption{\textbf{(a-c)} 2D projections of the 3D training set vectors $\{\log(\textit{A}), \log(\Delta), \log(\nabla)\}$
for the BPs (``$\diamond$'') and faculae (``$\triangle$'') classes of the training set, together with an axis
indicating the projected direction of maximum separability $\widehat{\bm{a}}$ (dashed-dot line). \textbf{d)}
Spline-interpolated density function histogram of the variable $F$, with the BP-faculae threshold $F_t$ (long
dashed line) and the rejection range corresponding to the apparent rejection rate $\alpha = 0.2$ (grey zone).}
\label{fig_LDA}
\end{figure*}

Having the correct parameters in hand, LDA could then be carried out to find their linear combination that best
discriminates the training set classes. Explicitely, we searched for the axis vector $\widehat{\bm{a}}$ that
maximizes Fischer's separability criterion \citep[as introduced in the original work of][]{Fischer36} in the
parameter space $\{\bm{x} = (\log(\textit{A}), \log(\Delta), \log(\nabla))\}$:
\begin{equation}\label{eqn_J}
\centering
J(\bm{a}) = \frac{[\bm{a}^{\rm T} (\bm{m}_{\rm bp} - \bm{m}_{\rm fac})]^2}{\bm{a}^{\rm T} (S_{\rm bp} + S_{\rm fac})
\bm{a}},
\end{equation}
where the superscript T denotes transpose, $\bm{m}_{\rm bp}$ and $\bm{m}_{\rm fac}$ are the class mean vectors
and $S_{\rm bp}, S_{\rm fac}$ the covariance matrices. The original parameters could then be projected onto
$\widehat{\bm{a}}$, thereby obtaining the desired linear combination defining the single variable $F \equiv
\widehat{\bm{a}}^{\rm T} \bm{x}$ (for ``Fischer'' variable). The 2D projections of the 3D training set vectors
are represented in Fig. \ref{fig_LDA}a-c, with an overlaid axis corresponding to the direction of maximum
separability $\widehat{\bm{a}}$. Fig. \ref{fig_LDA}d displays the DF of the obtained variable $F$. The maximal
value of $J$ associated with the variable $F$ (projection onto $\widehat{\bm{a}}$) and the values of $J$
associated to each parameter (projection onto the parameters' axes) are listed in Table \ref{table_params}. These
values give an idea of the relative ``discriminant power'' of the three parameters, and the larger $J$ value of
the variable $F$ demonstrates the advantage of their optimal linear combination provided by LDA.

The DFs of the discriminant parameters (Fig. \ref{fig_params} right column) and of the variable $F$ (Fig.
\ref{fig_LDA}d) also give a good visual estimate of the amount of overlap between the training set classes. An
intuitive measure of the discriminant power of a parameter could then be given by the ratio between the number of
features contained in the overlapping part of the DFs and the total number of training
set features. This ratio already takes a fairly low value of 0.07 for log($\Delta$), and goes down to 0.042 for
$F$. However, such a measure is statistically poor compared to $J$, since it mostly relies on the outliers
contained in the tails of the DFs (only 28 and 17 features for log($\Delta$) and $F$, respectively), whereas $J$
takes advantage of the full parameter distributions.

The overlap of the DF of log($\textit{A}$) and the particular skewness of the DF of faculae towards small
areas arises from our MLT segmentation. To investigate the influence of the MLT levels on the DFs of the
dicriminant parameters and on LDA, we carried out tests with fewer MLT levels over the same training set. We
found that the skewness of the DF of log($\textit{A}$) for faculae is in major part due to their segmentation
into fine striations. It should be noticed that the DF of log($\Delta$) is less skewed, because the
characteristic profiles of these striated faculae are mostly retrieved along the long dimension of the striations
(owing to their individual orientation, see Sect. \ref{sec_profiles}), which makes $\Delta$ a robust parameter to
distinguish them from BPs. However, the coarser segmentation of the tests has the undesired effect that a part of
the features are undersegmented, which leads to lower values of $J$ for all parameters as well as for the
discriminant variable $F$. Due to the merging of BPs into chains and ribbons, their DFs are particularly affected
and become skewed towards larger $\textit{A}, \Delta$ and smaller $\nabla$. For $\Delta$ and $\nabla$, this is
probably a consequence of the misorientation of merged features when retrieving the characteristic profiles. We
believe that those tests confirm our appropriate choice of MLT levels for the purpose of further discriminating
between individual BPs and faculae.

We stress that the procedure of orienting the features prior to the retrieval of their contrast profiles,
as described in Sect. \ref{sec_profiles}, is an essential ingredient to obtain discriminant parameters based
on those profiles. In early stages of this work, profiles were only retrieved along the direction
perpendicular to the closest limb, and the ensuing overlap of the DFs was significantly larger.

Finally, it should be noted that the values of these photometric discriminant parameters all depend to some extent
on the spatial resolution. $\nabla$ is probably the most sensitive in that respect, but it has the least weight
in the variable $F$ due to its lower value of $J$ (see Table \ref{table_params}). This points to the
requirement of having a dataset of roughly constant resolution, a condition met by our selection of images (Sect.
\ref{sec_processing}).

\begin{table}
\caption{$J$ values associated with each discriminant parameter, and for the variable $F$ obtained by linear
combination of the three parameters. The $a$'s are the coefficients of the linear combination (absolute values).}
\label{table_params}
\centering
\begin{tabular}{c c c c c c}
\hline\hline
       & \vline & log($\textit{A}$) & log($\Delta$) & log($\nabla$) & $F$(3D)\\    % table heading
\hline                        % inserts single horizontal line
   $J$ & \vline & 3.17 & 4.71 & 1.8 & 6.27 \\      % inserting body of the table
\hline                                   %inserts single line
   $a$ & \vline & 0.27 & 0.59 & 0.14 &  ...    \\      % inserting body of the table
\hline
\end{tabular}
\end{table}

\section{Classification results and discussion}\label{sec_classification}

\subsection{Hard threshold vs. reject option}\label{sec_rtvsro}

To build an assignment rule, we made the usual choice of a threshold value $F_{\rm t}$ at equal ``standardized''
distance from the class means \citep{Maha36, Murtagh87}, namely:
\begin{equation}\label{eqn_Ft}
\frac{(\widehat{\bm{a}}^T \bm{m}_{\rm bp} - F_{\rm t})^2}{\widehat{\bm{a}}^T
S_{\rm bp}\widehat{\bm{a}}} = \frac{(\widehat{\bm{a}}^T \bm{m}_{\rm fac} - F_{\rm t})^2 }{\widehat{\bm{a}}^T
S_{\rm fac}\widehat{\bm{a}}}.
\end{equation}
This threshold is drawn on the density function histogram of $F$ in Fig. \ref{fig_LDA}d.

\begin{figure*}
\centering
\includegraphics[width=\textwidth]{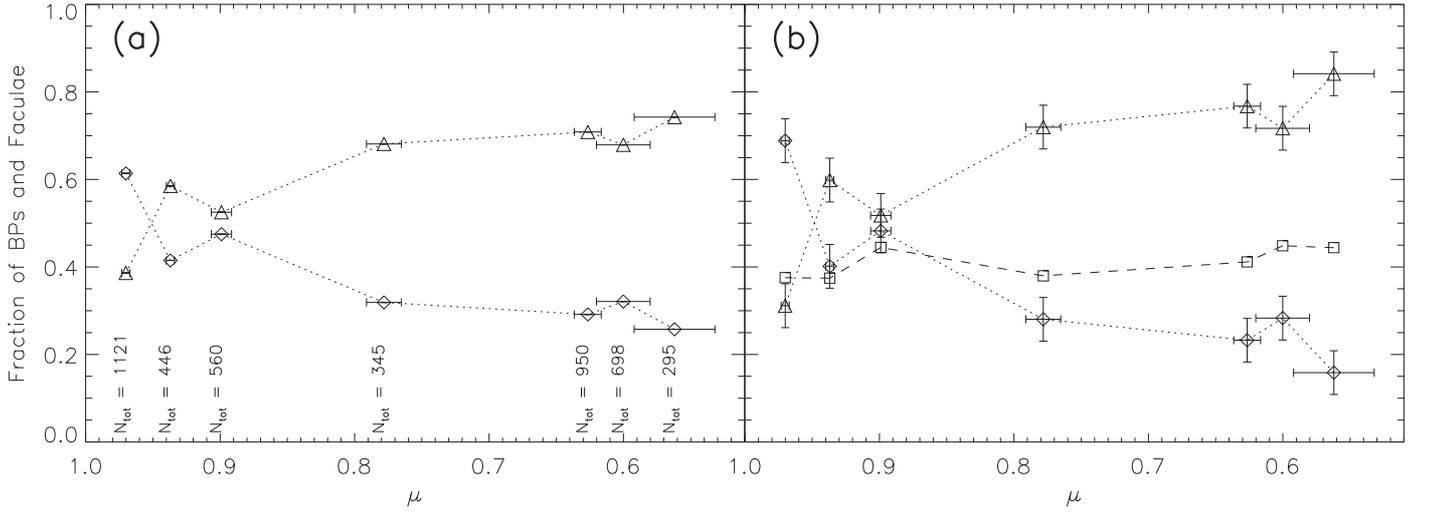}
\caption{\textbf{a)} CLV of relative fractions of BPs (``$\diamond$'') and faculae (``$\triangle$'') obtained by classifying all the segmented features
using a hard threshold $F_t$ defined via equation (\ref{eqn_Ft}). The abcissa of the points are the
$\left<\mu\right>$ of the analyzed images, and the horizontal error bars correspond to the standard deviation of
the individual features' $\mu$ values with respect to $\left<\mu\right>$. $N_{\rm tot}$ gives the total number of
features at each disk position, which depends on the effective FOV of the images and the number of available
image pairs at each disk position (see Table \ref{table_data}). The dotted lines are guides. \textbf{b)} Relative fractions of BPs (``$\diamond$'') and faculae (``$\triangle$'') after the introduction of a rejection range such that the apparent
rejection rates of BPs and faculae are equal to 0.2. The vertical error bars ($\pm 5 \%$) are upper limit estimates of the
errors induced by uneven true rejections rates (see main text for details). The fraction of rejected features with respect to the total
number of features is also plotted, as squares (``$\Box$'') joined by a dashed line.}
\label{fig_fractionCLV}
\end{figure*}

In a first step, a single hard threshold equal to $F_{\rm t}$ was used to classify \emph{all} the segmented
features as BPs or faculae by measuring their values of $F$. Because we are not interested in absolute numbers,
but rather in the relative distribution of BPs and faculae, we define the classified fractions of BPs and faculae as:
\begin{equation}\label{eqn_frac}
X_{\rm bp}   \equiv  \frac{N_{\rm bp}}{N_{\rm bp} + N_{\rm fac}}  , \quad
X_{\rm fac}   \equiv  \frac{N_{\rm fac}}{N_{\rm bp} + N_{\rm fac}},
\end{equation}
where $N_{\rm bp}$ and $N_{\rm fac}$ are the number of classified BPs and faculae, respectively. The CLV of the
fractions $X_{\rm bp}$ and $X_{\rm fac}$, classified using the threshold $F_{\rm t}$, is depicted in Fig.
\ref{fig_fractionCLV}a.

As already stated at the end of Sect. \ref{sec_scheme}, there is a continuous spectrum of intermediate
features between BPs and faculae that cannot be reasonably identified as belonging to one class or the other.
The only way to avoid the erroneous classification of these features is to exclude them from the statistics by
introducing a so-called ``reject option'' \citep{Hand81}, in the form of a rejection range in $F$ centered about
$F_{\rm t}$. Assuming that all intermediate features fall within the rejection range, the relations between
classified and true numbers ${N_{\rm bp}}^\ast$, ${N_{\rm fac}}^\ast$ (such as would be recognized by eye) at
each disk position are:
\begin{eqnarray}\label{eqn_num_rel_rej1}
N_{\rm bp}  & = & {N_{\rm bp}}^\ast (1-{\beta}^\ast)+ \epsilon^\ast {N_{\rm fac}}^\ast - \xi^\ast {N_{\rm bp}}^\ast, \\
N_{\rm fac} & = & {N_{\rm fac}}^\ast (1-{\alpha}^\ast)+ \xi^\ast {N_{\rm bp}}^\ast - \epsilon^\ast {N_{\rm
fac}}^\ast,
\end{eqnarray}
\begin{eqnarray}\label{eqn_num_rel_rej2}
N_{\rm rej} & = & {\alpha}^\ast{N_{\rm fac}}^\ast + {\beta}^\ast{N_{\rm bp}}^\ast + N_{\rm int}, \\
N_{\rm tot} & = & N_{\rm bp} + N_{\rm fac} + N_{\rm rej} = {N_{\rm bp}}^\ast + {N_{\rm fac}}^\ast + N_{\rm int},
\end{eqnarray}
where ${\alpha}^\ast$ and ${\beta}^\ast$ stand for the ``true'' rejection rates, i.e. the fractions of the actual
faculae and BP populations that fall in the rejection range, and $\epsilon^\ast$, $\xi^\ast$, represent the ``true''
misclassification rates. The boundaries of our rejection range were tuned to have equal ``apparent'' rejection
rates, $\alpha = \beta$, defined as the fractions of the \emph{training set} contained in that range. Assuming
that the training set adequately represents the true BPs and faculae populations, namely $\alpha^\ast \sim
\alpha$ and $\beta^\ast \sim \beta$, this precaution equalizes the true rejections as well ($\alpha^\ast \sim
\beta^\ast$), and thus prevents the introduction of bias in the number statistics. The apparent rejection rate
was set to a reasonable value for the rejection of intermediate features, $\alpha = 0.2$ (see below), yielding
the shaded rejection range on the histogram of $F$ in Fig. \ref{fig_LDA}d.

Under the full rejection of intermediate features, the classified fractions then become:
\begin{equation}\label{eqn_fracbp}
X_{\rm bp}   =  \frac{{N_{\rm bp}}^\ast (1-{\beta}^\ast)+ \epsilon^\ast {N_{\rm fac}}^\ast - \xi {N_{\rm bp}}^\ast}
{{N_{\rm bp}}^\ast (1-{\beta}^\ast) + {N_{\rm fac}}^\ast (1-{\alpha}^\ast)}
\end{equation}
for BPs and similarly for faculae. Note that if the training set is representative and the misclassification
rates can be considered negligible, relation \ref{eqn_fracbp} further simplifies to $X_{\rm bp} \sim X_{\rm
bp}^\ast$, where $X_{\rm bp}^\ast$ stands for the true fraction of BPs, namely $N_{\rm bp}^\ast/(N_{\rm bp}^\ast + N_{\rm
fac}^\ast)$ (same remark for faculae). The CLV of classified fractions $X_{\rm bp}$ and $X_{\rm fac}$, obtained
by adding a rejection range with $\alpha$ = 0.2, is shown in Fig. \ref{fig_fractionCLV}b.

Compared with the CLV of $X_{\rm bp}$ and $X_{\rm fac}$ obtained with a hard threshold, the difference
between $X_{\rm bp}$ and $X_{\rm fac}$ at each $\left<\mu\right>$ is now systematically larger (except for
$\left<\mu\right> = 0.9$). This is probably an effect of the contamination by intermediate features in the hard
threshold case, because such features are assigned roughly equally to each class, so that they have the tendency to
equalize $X_{\rm bp}$ and $X_{\rm fac}$. The difference between $X_{\rm bp}$ and $X_{\rm fac}$ is particularly
large for $\left<\mu\right> = 0.97$ and for the limbward data points at $\left<\mu\right> \leq 0.64$. This is
likely to be attributed to the larger misclassification errors in the hard threshold method than with the reject
option. Indeed, as can be seen from relation (\ref{eqn_num_rel_rej1}), the misclassification errors also tend to
overestimate the number of BPs near the limb where faculae dominate, while underestimating it near disk center,
and \textit{vice versa} for faculae. To estimate the true misclassification rates, we computed the apparent
misclassification rates $\epsilon, \xi$ by reclassifying the training set, assuming as before that the latter
adequately represents the true populations. With the chosen rejection range, we obtained $\epsilon, \xi = 0.005$,
which is an order of magnitude smaller than in the the hard threshold case and should be reflected in the true
rates as well\footnote{As these are only optimistic estimates fo the true misclassification rates, we implemented a
``leave-one-out'' method on the training set \citep{Hand81, Murtagh87}, which nevertheless gave the same results
as the simple reclassification (probably due to the fairly large size of the training set).}.

We shall now elaborate on the validity of the aforementioned assumptions, and further justify the use of
a reject option as opposed to the hard threshold classification. A subtle source of error is the
departure of the actual populations from the training set ones, causing $\alpha^\ast \neq \beta^\ast$. To evaluate
the importance of this effect, we have varied $\alpha$ in the range (0.2, 0.5), which should in that case induce unequal
variation of $\alpha^\ast$ and $\beta^\ast$ and consequently different variations of $X_{\rm bp}$ and $X_{\rm
fac}$ (cf. Eq. \ref{eqn_fracbp}). By the same token, this allowed us to check if intermediate features
were still wrongly classified as BPs or faculae for $\alpha = 0.2$, as the separation between $X_{\rm bp}$ and
$X_{\rm fac}$ should then increase with $\alpha$. But instead, we observed both positive and negative
fluctuations of $X_{\rm bp}$ and $X_{\rm fac}$, indicating that most intermediate features were indeed rejected,
and that the true rejection rates were nearly equal for BPs and faculae. As those fluctuations were always less
than 0.05, we chose this value as an upper limit on the error induced by uneven actual rejection rates, and
represented it by the symmetric error bars in Fig. \ref{fig_fractionCLV}b. To compare the effect of rejection at
the various disk positions, we overplotted in Fig. \ref{fig_fractionCLV}b the fraction of rejected
features with respect to the total number of features $N_{\rm rej}/N_{\rm tot}$. The relative constancy of
$N_{\rm rej}/N_{\rm tot}$ is reassuring, and reflects the self-similarity of the actual BPs and faculae
populations at various $\left<\mu\right>$ (as far as $F$ is concerned). It also gives an indication about the
number of intermediate features, as in absence of them we would have $N_{\rm rej}/N_{\rm tot} \simeq \alpha$,
using as before $\alpha^\ast \sim \beta^\ast \sim \alpha$ together with the relation (\ref{eqn_num_rel_rej2}). As
can be seen, the fraction of rejected features fluctuates around 0.4, indicating a significant fraction of
intermediate features $N_{\rm int}/N_{\rm tot} \simeq 0.2$, which further justifies the introduction of the
rejection range.

From a methodological point of view, despite the qualitative resemblance of the CLVs obtained with the
hard threshold and with the reject option, the first method is open to criticism due to the large amount of
intermediate features. A hard threshold is in this sense self-contradictory, as it tries to assign these features
to classes whose reference (the training set) does not represent them. By contrast, if these features are
properly rejected, and if the assumptions of negligible misclassification rates and representative training set
($\alpha^\ast \sim \beta^\ast$) are justified, the reject option method has the elegance that the classified
fractions closely approximate the true ones.

Lastly, the results obtained here do not only depend on the choice of the training set, but also on the choice of
the classification method. Fischer's LDA implicitely assumes similar covariance matrices for the classes, which
is not quite true in our case, as can be seen from the different shapes of the BPs and faculae ``clouds'' in
their 2D projections (Fig. \ref{fig_LDA}). We then implemented a ``class-dependent'' LDA, taking into account the
difference in covariance matrices and deriving different discriminant axes for the two classes \citep{Bala99}.
However, the difference in the relative fractions obtained was insignificant, thereby indicating that the
covariance matrices of our chosen parameters were suitable for Fischer's LDA.

\subsection{Discussion}\label{sec_discussion}

%Comparison with observations of faculae at DC and Bps off DC: confirmation of our method
To assess the validity of the proposed method, the obtained results can be compared with the observations of BPs
and faculae at various $\mu$ available in the literature. To help the comparison as well as to give a visual
idea of which features were classified as BPs and faculae, Fig. \ref{fig_mu97contours}, Fig.
\ref{fig_mu6contours} and Fig. \ref{fig_mu9contours} show their contours overlaid on the G-band images at
$\left<\mu\right> = 0.97$, $\left<\mu\right> = 0.6$ (same images as in Figs. \ref{fig_imagemu97} and
\ref{fig_imagemu6}), and $\left<\mu\right> = 0.9$ respectively. Our results are consistent with recent
high-resolution observations from \citet{Berger07}, who noticed the presence of disk-center faculae at
$\left<\mu\right> = 0.97$, very few intergranular BPs at $\left<\mu\right> = 0.6$, and a mixture of both features
at $\left<\mu\right> = 0.89$. The presence of some ``intergranular brightenings'' around $\left<\mu\right> =
0.55$ also has been reported by \citet{Lites04}, and \citet{Hirz05} have clearly observed the coexistence of BPs
and faculae at $\mu \sim 0.78$. This suggests the validity of our classification method, although this should be
confirmed in the future by using datasets with co-temporal magnetic vector information.

%Transition Bps to faculae
Although our CLV cannot be generalized due to the limited statistics of the present dataset and the coarse
sampling of the $\mu$ range, it allows us to constrain the $\mu$ interval where the transition from BPs to faculae
occurs. In this respect, the CLV also exhibits a plateau in the range $0.6 < \mu <0.78$, indicating that
BPs may still be found in that range, but progressively disappear closer to the limb, probably affected by the
foreground granular obscuration \citep{Auffret91}. This plateau can also be attributed to the slower variation of
the heliocentric angle in that $\mu$ range ($36^\circ < \theta < 54^\circ$) compared to the centerward range
$0.78 < \mu < 0.97$ ($13^\circ < \theta < 41^\circ$). Conversely, faculae appear to be present at all disk
positions, except for the inner third of the disk where $\mu > 0.9$. Hence, in contrast to full-disk images in
which faculae patches are only prominent closer to the limb, at high resolution faculae are conspicuous features
of active-region plages at all disk positions.

%Discussion of CLV in terms of inclined fiedls
The overall dominance of faculae in our dataset as well as the presence of BPs at relatively small $\mu$ values
($\mu \sim 0.6$) cannot be understood in terms of the conventional ``hot wall'' picture, if we consider only
vertical flux tubes and varying viewing angles with disk position. The most straightforward alternative is to
invoke inclined fields (e.g. due to swaying motions), whereby BPs would arise from flux tubes aligned along the
line of sight and faculae from flux tubes inclined with respect to it. Again, such an hypothesis should be
verified with the help of magnetic vector data.

%Assets and Pitfalls
From here on, we discuss the assets and weaknesses of the proposed method, as well as its
applicability. A key point of the method resides in the orientation of individually segmented features so as to
retrieve characteristic profiles (see Sect. \ref{sec_profiles}). This procedure makes the method applicable at
different $\left<\mu\right>$ (due to the diverse orientations of BPs and faculae), and thereby offers the
possibility of studying the transition from BPs to faculae as $\mu$ varies. In the considered $\mu$ range at
least, the orientation process makes the discriminant parameters roughly $\mu$-invariant, thus allowing LDA to be
applied to the whole dataset at once (cf. Sect. \ref{sec_params}). LDA itself has the advantage of being fairly
simple to implement, and of making few assumptions on the distribution properties of the discriminant
parameters (see Sect. \ref{sec_LDA}). It nevertheless requires the careful preselection of a training set, a
crucial step that can potentially bias the classification. But the principal weakness of the current method lies
in the use of photometric information only, allowing a limited number of discriminant parameters to be
defined. This induces the following pitfall: faculae ``sitting'' on very small granules (fragments, abnormal
granulation) are basically indistinguishible from BPs as far as our parameters are concerned. Several instances
appear in Figs. \ref{fig_mu6contours} and \ref{fig_mu9contours}, where such small faculae are either rejected or
misclassified. The method could be improved by the inclusion of discriminant parameters coming from polarimetric
maps, provided that BPs and faculae exhibit sufficiently different magnetic properties.

%Applicability
We conclude by drawing attention to precautions that should be taken in applying our method to other datasets.
Having a fairly homogeneous and high spatial resolution is an essential requirement, as the values of all
parameters $\textit{A}$, $\Delta$ and $\nabla$ depend on it. A variable resolution would cause the values of the
parameters to vary throughout the dataset (between different images or even accross the field of view), thus
preventing a well-defined discriminant variable and a unique BP/faculae threshold from being obtained. For a
dataset with a constant but different resolution, the method would in principle still be valid, but the values of
the parameters and of the class threshold would differ. However, degrading datasets of variable resolution to a
constant lower one would reduce the contrast of features (loss of statistics due to the contrast threshold), and
blur the local contrast depressions, so that it would become difficult to separate adjacent BPs and faculae
striations. The method would also lose in efficiency due to the misorientation of merged features (cf. Sect.
\ref{sec_profiles}).
Finally, care
should be taken in applying unchanged the herein-derived discriminant $F$ and its threshold value to other
datasets. If the current method is applied to a dataset of slightly different resolution (or with a different
amount of straylight), wavelengths \citep[e.g. CN-band, ][]{Zakharov07} or $\mu$ range, the values of the
segmentation thresholds should first be adapted (the same holds for the identification criteria of the
``grey-lane'' BPs), and the training set selection must then be repeated (as well as the subsequent steps of the
method thereof). Under different conditions, the ensuing values of the discriminant parameters will be different,
and consequently LDA will yield a different linear combination for the discriminant variable and a different
threshold.

%______________________________________________________________

\section{Summary and outlook}\label{sec_summary}

We have developed a photometric method based on Linear Discriminant Analysis (LDA) to discriminate
between individual Bright Points (BPs) and faculae, observed at high resolution over a range of heliocentric
angles. We first demonstrated the feasibility of an automated segmentation for both individual BPs and faculae
at various disk positions, based on joint G-band and continuum photometric information only. For each segmented
feature, a ``characteristic G-band contrast profile'' was retrieved along a specific direction, by properly
orienting the feature using its ``contrast moment of inertia''. Three physical parameters were then identified to
be good discriminants between BPs and faculae at all disk positions of our dataset: the width and slope of the
contrast profiles, as well as the apparent area defined by the segmentation map. Linear discriminant
analysis was then performed on a visually-selected reference set of BPs and faculae, yielding a single linear
combination of the parameters as the discriminant variable for all disk positions. Using an appropriate threshold and
rejection range on this variable, all the segmented features were ultimately classified and the relative
fractions of BPs and faculae at each disk position of our dataset were computed. The resulting CLV of these
fractions is mostly faculae-dominated except for $\mu > 0.9$, i.e. close to disk center. This is in agreement
with previous observations, thus suggesting the validity of the presented method. We propose that these
ubiquitous faculae are produced by a hot-wall effect through \emph{inclined fields}.

Using our classification method, we plan to present more statistical results concerning photometric properties of
BPs and faculae (such as contrast and morphology) in a forthcoming paper. A similar classification study should
in future also be performed on a high-resolution dataset with magnetic field vector information, in order to
determine the magnetic properties of BPs and faculae separately and further validate the method. Such
datasets can now be obtained from ground-based imaging polarimeters such as the GFPI \citep{GFPI}, CRISP
\citep{CRISP}, or IBIS \citep{IBIS}, and in the near future from the SUNRISE \citep{SUNRISE} stratospheric
balloon-born observatory. Through their seeing-free quality, the images of SUNRISE will be very
promising for the application of our method, as they will naturally satisfy its requirement of homogeneous
spatial resolution. In particular, a comparison of the classification results with the field inclinations
retrieved by Stokes profile inversions could be particularly interesting to verify the hypothetical association
of faculae with inclined fields. Ultimately, a similar classification should be performed on synthetic images
computed from 3D MHD simulation boxes at various angles. A comparison with the classification obtained from
observational data would then give more physical insight into the relationship between BPs and faculae on one hand,
and provide novel constraints for the models on the other hand.

%Figures 11-14 are only available online

\onlfig{11}{
\begin{figure*}
\centering
\includegraphics[width=\textwidth]{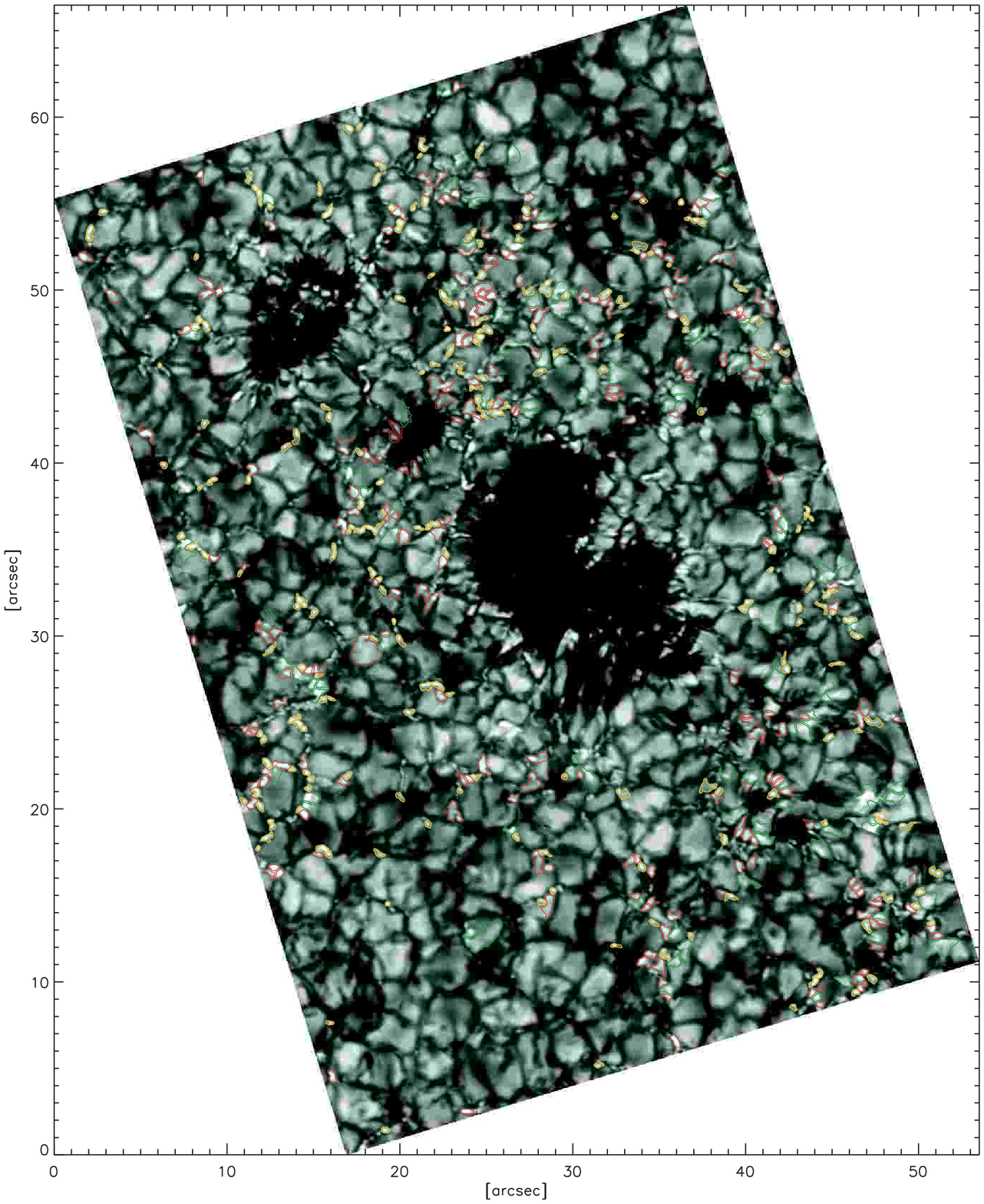}
\caption{G-band image at $\left<\mu\right> = 0.97$ (the same as in Fig. \ref{fig_imagemu97}), with the overlaid
contours of the features classified using the reject option. The faculae are contoured in green, the BPs in
yellow and the rejected features in red. The contours correspond to the border of the features as defined by the
segmentation map (corresponding to the lowest MLT level $C_{\rm G} = 0$). Tickmarks are in arcseconds. The
corresponding original image without contours is presented in Fig. \ref{fig_imagemu97}.}
\label{fig_mu97contours}
\end{figure*}
}

\onlfig{12}{
\begin{figure*}
\centering
\includegraphics[width=\textwidth]{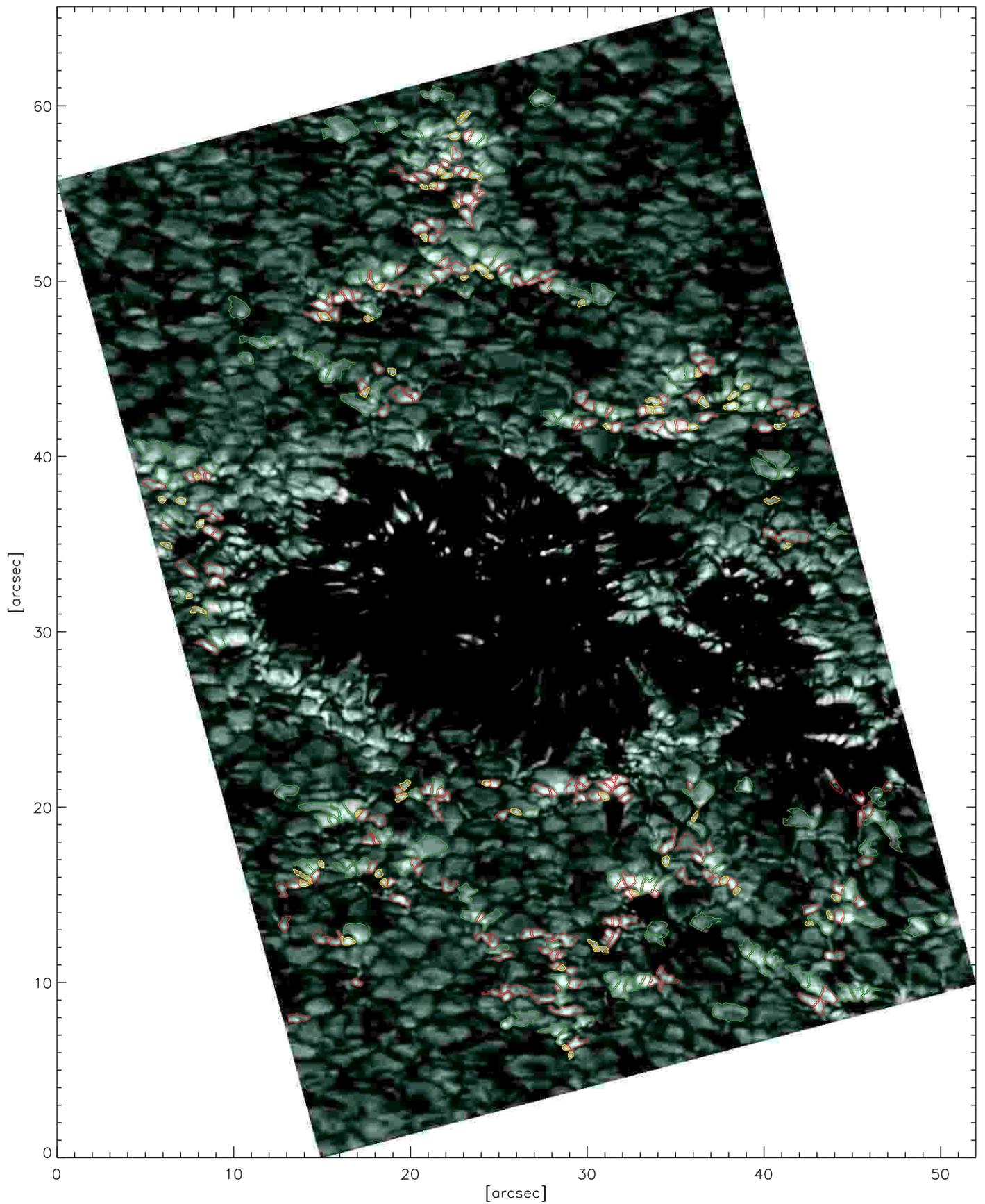}
\caption{G-band image at $\left<\mu\right> = 0.6$ (the same as in Fig. \ref{fig_imagemu6}), with the overlaid
contours of the features classified using the reject option (see caption of Fig. \ref{fig_mu97contours}). The
corresponding original image without contours is presented in Fig. \ref{fig_imagemu6}.}
\label{fig_mu6contours}
\end{figure*}
}

\onlfig{13}{
\begin{figure*}
\centering
\includegraphics[width=\textwidth]{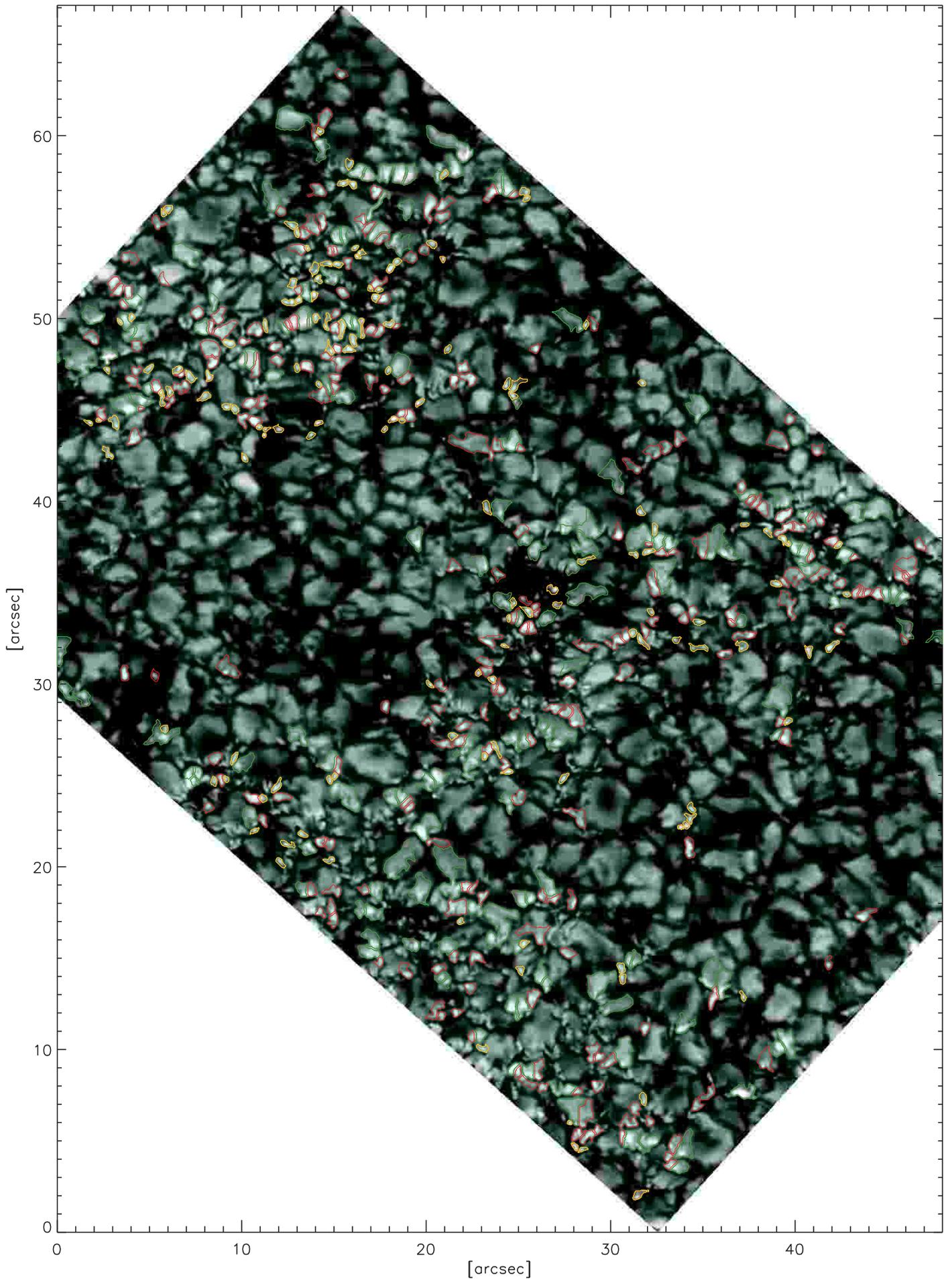}
\caption{G-band image at $\left<\mu\right> = 0.9$, with the overlaid contours of the features classified using
the reject option (see caption of Fig. \ref{fig_mu97contours}). The original field of view has been slightly
cropped horizontally for this display. The direction of the closest limb is upwards.}
\label{fig_mu9contours}
\end{figure*}
}

\onlfig{14}{
\begin{figure*}
\centering
\includegraphics[width=\textwidth]{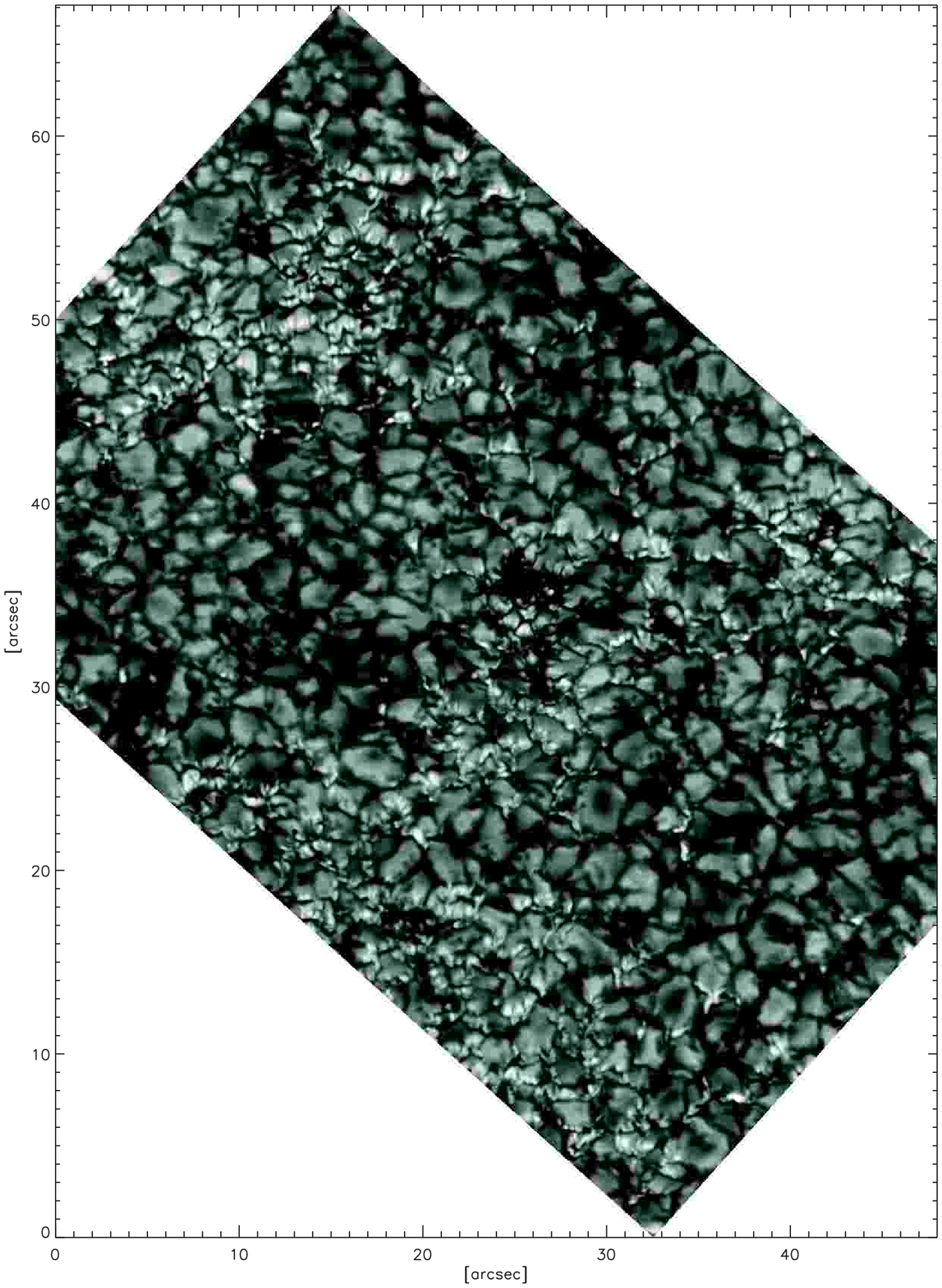}
\caption{Original G-band image at $\left<\mu\right> = 0.9$ without overlaid contours, for comparison with Fig. \ref{fig_mu9contours}.}
\label{fig_imagemu9}
\end{figure*}
}

\begin{acknowledgements}
We are grateful to F. Kneer, E. Wiehr, O. Steiner, F. Murtagh, M. Sch{\"u}ssler and W. Stahel for their interest
in the present work and their fruitful comments. We thank as well P. S{\"u}tterlin for having provided the basis
of our image warping code. This work was partly supported by the WCU grant No. R31-10016 from the Korean Ministry
of Education, Science and Technology. Finally, our gratitude is directed towards D.~Schmitt and the ``Solar
System School'', in which frame this research could be carried out.
\end{acknowledgements}

%Using BibteX___________________________________________________________
\bibliographystyle{aa}
\bibliography{1117bib}

\end{document}